\documentclass[
    prl, twocolumn, superscriptaddress, nofootinbib, amsmath, amssymb,
    aps, floatfix, preprintnumbers
]{revtex4-2}
\usepackage[utf8]{inputenc}
\usepackage{setspace}
\usepackage[dvips]{graphicx}
\usepackage[dvipsnames]{xcolor}
\definecolor{CiteBlue}{RGB}{45,52,151}
\usepackage[
    colorlinks=true,
    linkcolor=CiteBlue,
    urlcolor=CiteBlue,
    citecolor=CiteBlue
]{hyperref}
\usepackage{aas_macros}
\usepackage[capitalise]{cleveref}
\usepackage[version=4]{mhchem}
\usepackage{siunitx}
\DeclareSIUnit{\year}{yr}

\newcommand{\refcite}[1]{Ref.~\cite{#1}}
\newcommand{\refscite}[1]{Refs.~\cite{#1}}

\usepackage{bm}
\newcommand{\bb}[1]{\bm{\mathrm{#1}}}
\newcommand{\du}{\mathrm{d}}
\newcommand{\dd}{\,\du}

\newcommand{\dm}{\chi}

\newcommand{\earth}{\mathrm{e}}
\newcommand{\esc}{\mathrm{esc}}
\newcommand{\phonon}{{\mathrm{ph}}}

\renewcommand{\subsection}[1]{\textit{#1.---}\ignorespaces}

\begin{document}
\title{
Dark Matter--Electron Detectors for Dark Matter--Nucleon Interactions}
\preprint{MIT-CTP/5678}

\author{Sin\'{e}ad M. Griffin}
\affiliation{Materials Sciences Division, Lawrence Berkeley National Laboratory, Berkeley, CA 94720, USA}
\affiliation{Molecular Foundry, Lawrence Berkeley National Laboratory, Berkeley, CA 94720, USA}

\author{Guy Daniel Hadas}
\affiliation{Racah Institute of Physics, Hebrew University of Jerusalem, Jerusalem 91904, Israel}

\author{Yonit Hochberg}
\affiliation{Racah Institute of Physics, Hebrew University of Jerusalem, Jerusalem 91904, Israel}
\affiliation{Laboratory for Elementary Particle Physics, Cornell University, Ithaca, NY 14853, USA}

\author{Katherine Inzani}
\affiliation{School of Chemistry, University of Nottingham, United Kingdom}

\author{Benjamin V. Lehmann}
\affiliation{Center for Theoretical Physics, Massachusetts Institute of Technology, Cambridge, MA 02139, USA}

\date\today

\begin{abstract}\ignorespaces{}
    In a seminal paper now a decade old, it was shown that dark matter detectors geared at probing interactions with nucleons could also be used to probe dark matter interactions with electrons. In this work, we show that new detector concepts designed to probe dark matter--electron interactions at low masses can similarly be used to probe new parameter space for dark matter--nucleon interactions. We demonstrate the power of this approach by using existing data from superconducting detectors to place new limits on the interactions of nuclei with MeV-scale dark matter. Further, we show that advances in detector technology that have been anticipated for electronic interactions will automatically extend sensitivity deep into uncharted territory for nuclear interactions. This doubles the effective science output of future low-threshold experiments.
\end{abstract}

\maketitle
\section{Introduction}
\label{sec:intro}
The nature of dark matter (DM) remains one of the biggest open problems in modern physics. For decades, most laboratory DM searches operated under the assumption that DM particles would be associated with the weak scale of the Standard Model, and thus targeted heavy DM particles with masses well above the GeV scale~\cite{Lee:1977ua,Kolb:1990vq,Jungman:1995df,Bergstrom:2000pn,Bertone:2004pz}. In the wake of null results, the community has shifted focus to a broader set of candidates, and has developed searches sensitive to different mass scales and interactions. One notable step of this transition was the realization by the authors of \refcite{Essig:2011nj} that existing experiments designed to search for DM-nucleon scattering could also be used to probe DM-electron scattering at masses well below the GeV scale. This set off a new exploratory phase in the development of DM searches, and experimental sensitivity to DM-electron interactions was quickly extended to DM masses well below those probed via nuclear recoils.  

Since then, advancements in technology have yielded many new detection modalities based on a variety of different electronic systems, offering sensitivity to extremely low energy deposits and thus to low DM masses~\cite{Essig:2011nj,Graham:2012su,Essig:2015cda,Hochberg:2015pha,Hochberg:2015fth,Hochberg:2019cyy,Hochberg:2021ymx,Hochberg:2021yud,Derenzo:2016fse,Hochberg:2016ntt,Hochberg:2017wce,Cavoto:2017otc,Kurinsky:2019pgb,Blanco:2019lrf,Griffin:2020lgd,Simchony:2024kcn,Essig:2022dfa,Das:2022srn,Das:2024jdz}. Among the proposed platforms, superconductors~\cite{Hochberg:2015pha, Hochberg:2015fth, Hochberg:2019cyy, Hochberg:2021ymx, Hochberg:2021yud} have been quickly advancing in both threshold and scale. The small gap of the superconducting phase, $\mathcal O(\qty{}{\milli\electronvolt})$ in common materials, allows in principle for the detection of DM down to the keV scale, where cosmological bounds become relevant. Several new experiments leveraging such new ideas are already underway~\cite{future:QROCODILE, future:TES}, and prototype superconducting detectors have already produced new limits on DM-electron interactions~\cite{Hochberg:2019cyy,Hochberg:2021yud,Gao:2024irf}.

However, a complete and cohesive low-mass DM detection program must include detection strategies for DM-nucleon scattering as well as DM-electron scattering. Many interesting models feature only DM-nucleon interactions, and these would be invisible to an experiment that is only sensitive to DM-electron interactions. It is especially desirable to identify single systems that can probe both nuclear and electronic interactions simultaneously. A large class of electron-recoil detectors have been proposed and designed with the specific aim of detecting DM interactions with electrons themselves. This is sensible from a historical perspective, since their reach into low DM masses was originally driven by kinematics: when the DM mass is below the GeV scale, and the relevant process for detection is two-body elastic scattering, interactions with electrons can transfer energy much more efficiently than interactions with heavier nuclei.

But as the DM mass decreases further, the interaction rate becomes dominated by many-body processes rather than two-body elastic scattering. In this regime, detectors are primarily sensitive to the excitation of collective modes, such as the production of quasiparticles and phonons, rather than the recoil of individual electrons and nuclei. Here, a central challenge for DM searches is to detect extremely low energy deposits, regardless of the mode that sourced them. This is exactly where recent generations of electron-recoil detectors truly excel. Moreover, since the different classes of collective excitations are coupled to one another and to single-particle excitations, a system that specializes in detecting excitations amongst the electrons should have automatic sensitivity to excitations amongst the nuclei, and vice versa.

Accordingly, in this work, rather than asking whether a nuclear recoil detector can automatically detect low-energy electronic recoils, as was done in the seminal work of \refcite{Essig:2011nj}, we ask whether an electron recoil detector can automatically detect low-energy nuclear recoils. We show that experiments that were designed to detect DM-electron interactions can in fact \emph{already} probe DM-nucleon interactions at low DM masses.
Using existing data from superconducting detectors, and focusing on interactions with a scalar mediator, we place new limits on DM-nucleon scattering, in parameter space not otherwise constrained by elastic scattering experiments. We thus demonstrate the potential for low-threshold DM experiments to double their science output.

\section{Nuclear scattering events in electronic detectors}
\label{sec:nuclear-scattering}
For the purposes of the present work, we specialize to superconducting detectors, although similar statements hold for other classes of low-threshold electron recoil detectors. Superconducting detectors are ultimately sensitive not to single electron recoils, but to changes in the properties of the superconducting phase in the detector. For example, this might be an increase in the phonon density, as in a transition-edge sensor (TES)~\cite{Schwemmbauer:2024rcr}; or an increase in the quasiparticle density, as in a kinetic inductance detector (KID)~\cite{Gao:2024irf}; or a complete transition of a portion of the detector from the superconducting phase to the normal metal phase, as in a superconducting nanowire single photon detector (SNSPD)~\cite{Hochberg:2019cyy,Hochberg:2021yud}.

In all superconductor DM searches to date, only DM-electron interactions have been considered as possible triggers of such phase changes, but each of these effects can also be caused by DM-nucleon interactions. Low-energy DM-nucleon scattering events lead directly to the production of phonons. Phonons can, in turn, downconvert by breaking Cooper pairs and producing free quasiparticles. If the initial deposit is large enough, the excess phonons and quasiparticles can prompt a transition to the normal metal state, just as with electronic recoils. Each of the detectable phenomena in a superconducting detector is thus sensitive to DM-nucleon scattering.

In general, the dependence of the event rate on the properties of the target system is determined by the dynamic structure factor, $S\left( \bb q, \omega\right)$, which measures the response of the system to a deposited momentum $\bb q$ and energy $\omega$. Working in natural units with $c=\hbar=1$, the DM scattering rate then takes the form
\begin{multline}
    \label{eq:rate-from-structure-factor}
    R = \frac{\pi \overline{\sigma}_n \rho_{\dm}}{
        \mu^2_{\dm n} \rho_{\mathrm{T}} m_{\dm}}
        \int \frac{\du^3\bb q}{(2 \pi)^3}\dd^3\bb v\dd\omega
            \Bigl[f_{\dm}(\bb v) \mathcal{F}^2_{\mathrm{med}}(\bb q)
    \\
        \times  S\left( \bb q, \omega\right)
            \delta\left(\omega -\omega_{\bb q} \right)\Bigr]\,,
\end{multline}
where $\rho_{\dm} = \qty{0.4}{\giga\electronvolt/\centi\meter^3}$ is the local DM density; $\rho_{\mathrm T}$ is the target density; $\mathcal{F}^2_{\mathrm{med}}(\bb q)$ is a form factor determined by the structure of the interaction, given by $1$ for a heavy mediator and $\left(q_0 / q \right)^2$ for a light mediator, where $q_0=m_\dm v_0$ for $v_0$ the mean DM velocity and $q \equiv \left|\bb q\right|$; $f_{\dm}(\bb v)$ is the DM velocity distribution; and $\omega_{\bb q} \equiv \bb v \cdot \bb q - \bb q^2/2 m_{\dm}$ is the energy deposited. Given a matrix element $\mathcal M_{\dm n}(q) = \mathcal M_0 \mathcal F_{\mathrm{med}}(q)$, we define a reference cross section $\overline\sigma_n = (\mu_{\dm n}^2/\pi)\left|\mathcal M_{\dm n}(q_0)\right|^2$  where $\mu_{\dm n}$ is the reduced mass.
We take $f_\dm$ to be a shifted Maxwell-Boltzmann distribution with $v_0 = \qty{230}{\kilo\meter/\second}$, Earth velocity $v_\earth = \qty{240}{\kilo\meter/\second}$ in the Galactic frame, and escape velocity $v_\esc = \qty{600}{\kilo\meter/\second}$.

As we are considering multiple interaction channels for the DM, it is important to consider the differential sensitivity of the experimental system to energy deposited in each channel. In particular, TESs~\cite{Schwemmbauer:2024rcr} and SNSPDs~\cite{Hochberg:2019cyy,Hochberg:2021yud} are nominally sensitive to heat, while KIDs~\cite{Gao:2024irf} are nominally sensitive to quasiparticles in the final state. Fortunately, for energy deposits $\omega$ in a superconductor that are much larger than the superconducting gap $2\Delta$, the partitioning of energy between phonons and quasiparticles in the final state is independent of $\omega$ and~$q$. The detector threshold is typically derived from calibration experiments using photon absorption, and for all of the real detectors we consider in this work, these calibration experiments are performed with $\omega \gg 2\Delta$. Thus, the determination of the threshold is subject to the same partitioning of energy that occurs for an electronic or nuclear scattering event. This means that there is no need to treat DM interactions differently from photon absorption for the purposes of determining the threshold.

Determining the sensitivity to DM-nucleon interactions is now reduced to the computation of the dynamic structure factor for each interaction process of interest. Here, the processes we consider are elastic nuclear recoils, single phonon production, and multiple phonon production. We now calculate $S(\bb q, \omega)$ for each of these channels.

\subsection{Elastic nuclear recoils}
In general, the interaction between a DM particle and a nucleon in the target material is complicated by the response of the nuclear lattice. However, if  the deposited energy $\omega$ is much larger than the maximal phonon energy $\omega_\phonon^{\max}$, the DM-nucleon interaction is decoupled from the lattice structure of the target. If the target is composed of several types of atoms, then the dynamic structure factor for a hadrophilic interaction is given by a weighted average of the contributions of each atom, as~\cite{Trickle:2019nya}
\begin{equation}
    \label{eq:structure-factor-elastic}
    S(\bb q, \omega) = \frac{2\pi\rho_{\mathrm{T}}}{\sum_{N} A_{N}} \sum_{N} \frac{A_N^3}{m_N}
    F_N(\bb q)
    \delta \left(\omega - \frac{\bb q^2}{2 m_N}\right)
    \,.
\end{equation}
Here $N$ indexes the nuclei in a unit cell; $m_N$ is the atomic mass; $A_N = m_N/\qty{}{u}$ is the atomic mass number; $f_n$ is the coupling to DM\@; and $F_N(\bb q)$ is the nuclear form factor.
We use the Helm form factor~\cite{Helm:1956zz}, $F_N (q) = [3 j_1 (q r_N)/(q r_N)] e^{-(qs)^2/2}$,  where $j_1$ denotes the spherical Bessel function of the first kind; $r_N \approx A_N^{1/3}\times\qty{1.14}{\femto\meter}$ is the effective nuclear radius; and $s$ is the nuclear skin thickness. We use $\{A_{\ce{W}}, A_{\ce{Si}}, A_{\ce{Ti}}, A_{\ce{N}}, A_{\ce{Al}}\} \approx \{183.85, 28.09, 47.87, 14.01, 26.98\}$, and we set $s = \qty{0.9}{\femto\meter}$ for all materials.

\subsection{Single phonons}
\label{subsec:single_phonon}\ignorespaces
On the other hand, for the smallest deposits, $\omega \ll \omega_\phonon^{\max}$, the relevant excitations are single phonons. The dynamic structure factor for the production of a single phonon is given by~\cite{Trickle:2019nya}
\begin{multline}
    \label{eq:structure-factor-single-phonon}
    S(\bb q, \omega) = \frac{\pi}{\Omega} \sum_{\nu}
    \frac{\delta(\omega-\omega_{\nu, \bb k})}{\omega_{\nu, \bb k}}
    \Bigg| \sum_j \frac{e^{- W_j(\bb q)}}{\sqrt{m_j}}
        e^{i \bb G \cdot \bb x_j^0} \\
    \times\left( \frac{f_{N_j}}{f_n} \right)
    F_{N_j}(q) \bb q \cdot \bb{\epsilon}^*_{\nu, \bb k, j} \Bigg|^2 \,.
\end{multline}
Here $\Omega$ is the volume of the primitive cell; $\nu$ indexes the phonon branches, and $j$ indexes atoms in the primitive cell; $\bb q = \bb k + \bb G$, with $\bb k$ lying in the first Brillouin zone and with $\bb G$ a reciprocal lattice vector; $\omega_{\nu, \bb k}$ is the phonon frequency; $m_j$ is the atomic mass; $\bb{\epsilon}_{\nu, \bb k, j}$ is the polarization vector; $F_{N_j}(q)$ is the nuclear form factor; and $W_j(\bb q)$ is the Debye-Waller factor, which describes the physics of the creation and annihilation of phonons from the vacuum. The latter is given by
\begin{equation}
    \label{eq:full-debye-waller-factor}
    W_j(\bb q) = \frac{1}{4 N_L m_j} \sum_{\nu} \sum_{\bb k \in \mathrm{1BZ}}
    \frac{\left|\bb q \cdot \bb\epsilon_{\nu, \bb k, j} \right|^2}
        {\omega_{\nu, \bb k}}\,,
\end{equation}
where $N_L$ is the number of primitive unit cells in the lattice, which give rise to $N_L$ discrete values of $\bb k$. In practice, we take the continuum limit, $N_L\to\infty$. The Debye-Waller factor suppresses the contribution of higher momentum transfers to the phonon production rate.

\subsection{Multiple phonons}
\label{subsec:multi_phonon}\ignorespaces
For intermediate deposits which are neither much smaller nor much larger than $\omega_\phonon^{\max}$, an interaction can produce several phonons in the material. The analysis in this case is more involved. In this work, we use the incoherent approximation as described in \refcite{Campbell-Deem:2022fqm}, where dynamic structure factor is written as 
\begin{multline}
    \label{eq:structure-factor-multiphonon-incoherent}
    S(q, \omega) \approx \frac{2 \pi}{\Omega}
    \sum_j \left( \frac{f_{N_j}}{f_n} \right)^2 e^{-2 W_j\left(q\right)}
    \sum_n \frac{1}{n!}\left(\frac{q^2}{2 m_j} \right)^n\\
    \times \left(\prod_{i=1}^n \int\du\omega_i\,
    \frac{D_j\left(\omega_i\right)}{\omega_i} \right)
        \delta \left(\sum_i \omega_i - \omega \right)\,,
\end{multline}
where the $n$th term is the contribution of the $n$-phonon final state, and we sum terms up to $n = 10$. Here $D_j$ is the partial density of states for the $j$th atom in the unit cell, given by
\begin{equation}
    \label{eq:partial-density-of-states}
    D_j(\omega) =
    \frac{1}{3N_L} \sum_{\nu} \sum_{\bb k \in \mathrm{1BZ}}\left|
        \bb{\epsilon}_{\nu, \bb k, j}
    \right|^2 \delta\left(\omega - \omega_{\nu, \bb k}\right)
    \,.
\end{equation}
In \cref{eq:structure-factor-multiphonon-incoherent}, $W_j$ denotes the partial Debye-Waller factor, which is given by $W_j\left(q\right) = (q^2/4 m_j) \int\du\omega'\,D_j\left(\omega'\right)/\omega'$. 

We note that the incoherent approximation is not expected to be valid for momentum transfers smaller than the size of the Brillouin zone, $q_{\mathrm{BZ}}$. The length scale corresponding to $q < q_{\mathrm{BZ}}$ extends over more than one unit cell, so coherent contributions from atoms in different unit cells become important. In this region, the scattering rate is again dominated by single phonon excitations. We thus omit the $n = 1$ term in the sum of \cref{eq:structure-factor-multiphonon-incoherent} and add in its place the accurate single phonon contribution of \cref{eq:structure-factor-single-phonon}. 
At fixed $\omega$, the sum of contributions from $n$-phonon final states in \cref{eq:structure-factor-multiphonon-incoherent} converges more slowly as $q$ grows due to the factor of $q^2/2 m_j$. For this reason, we follow \refcite{Campbell-Deem:2022fqm} and use the impulse approximation for the dynamic structure factor when $q > 2 \sqrt{2 m_j \overline{\omega}_j}$, where $\overline{\omega}_j$ is the mean frequency weighted by the partial density of states $D_j$. The dynamic structure factor in the impulse approximation is then given by 
\begin{equation}
    \label{eq:structure-factor-impulse-approximation}
    S(q, \omega) = \frac{2 \pi}{\Omega} \sum_j
    \frac{\left( f_{N_j}/f_n \right)^2}{\sqrt{2 \pi \Delta_j^2}} \exp\left(
        - \frac{\left(\omega - q^2/2 m_j \right)^2}{2 \Delta_j^2}
    \right)
    \,,
\end{equation}
where the Gaussian width is $\Delta_j^2 = q^2 \overline{\omega}_j/(2 m_j)$.

\begin{figure}\centering
    \includegraphics[width=\columnwidth]{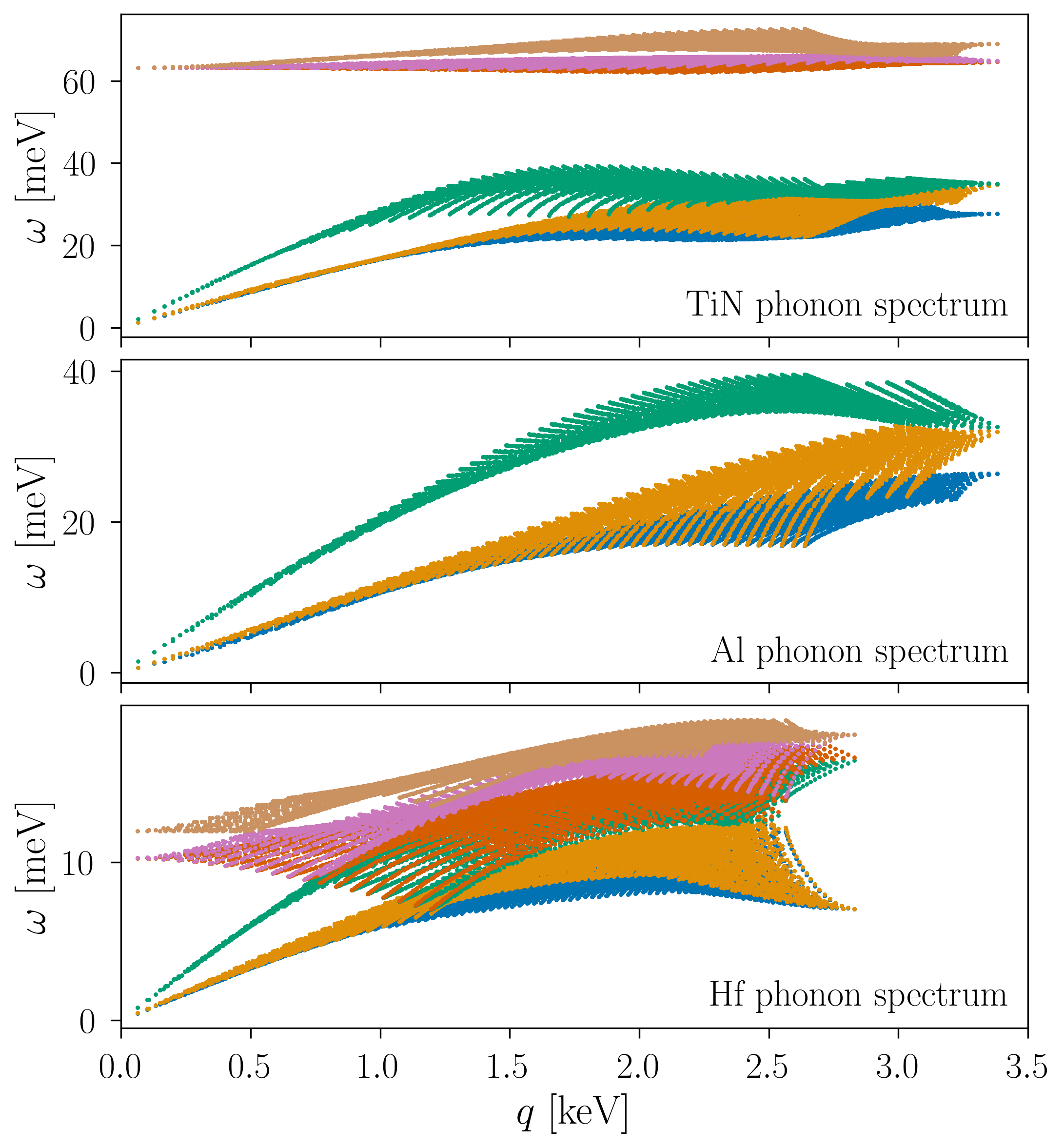}
    \caption{\textbf{Phonon spectra.} Computed spectrum of phonons in \ce{TiN} (\textit{top}), \ce{Al} (\textit{middle}), and \ce{Hf} (\textit{bottom}), with bands differentiated by color. \ce{TiN} and \ce{Hf} exhibit both acoustic phonons, which have linear dispersion at small $q$, and optical phonons, which have a nonzero gap with nearly flat dispersion. Spectra are computed in normal metal phase, as corrections from superconductivity are negligible at the deposits we consider.
    }
    \label{fig:phonon-spectrum}
\end{figure}

\section{Results}\label{sec:results}

\begin{figure*}\centering
    \includegraphics[width=0.495\textwidth]{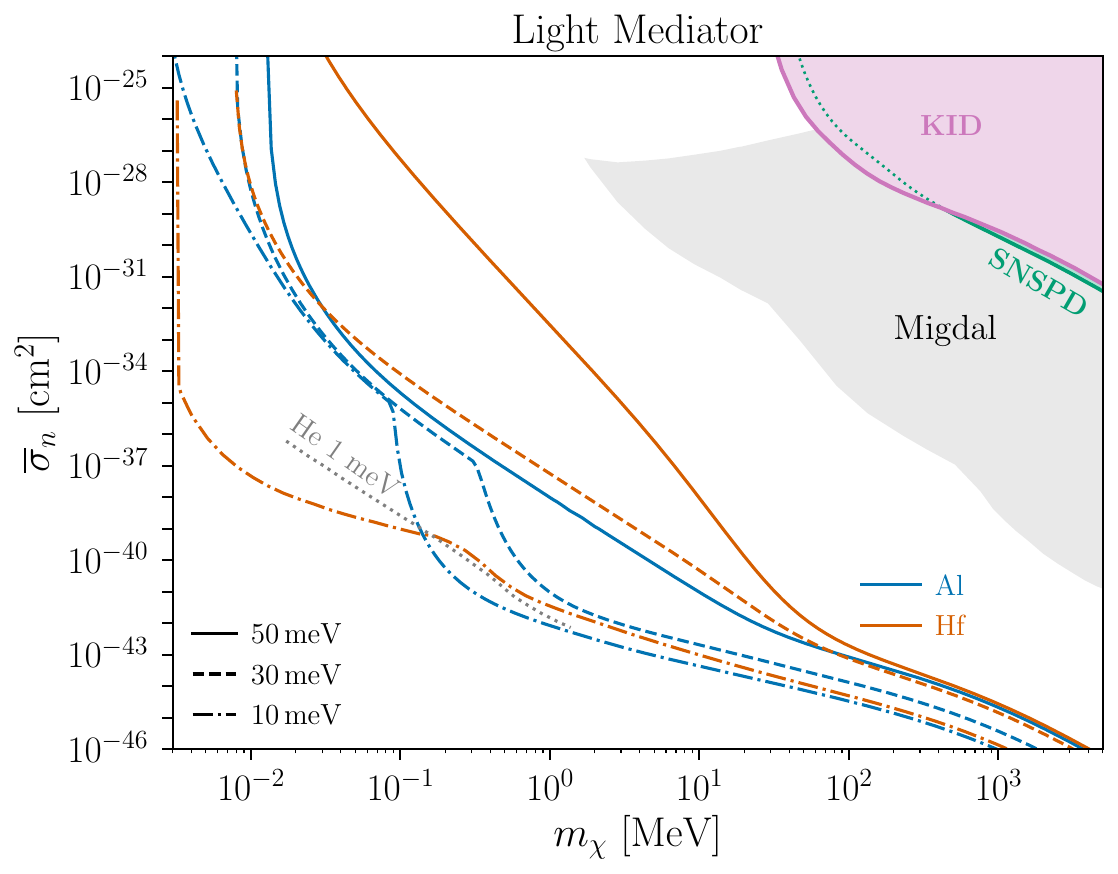}
    \hfill
    \includegraphics[width=0.495\textwidth]{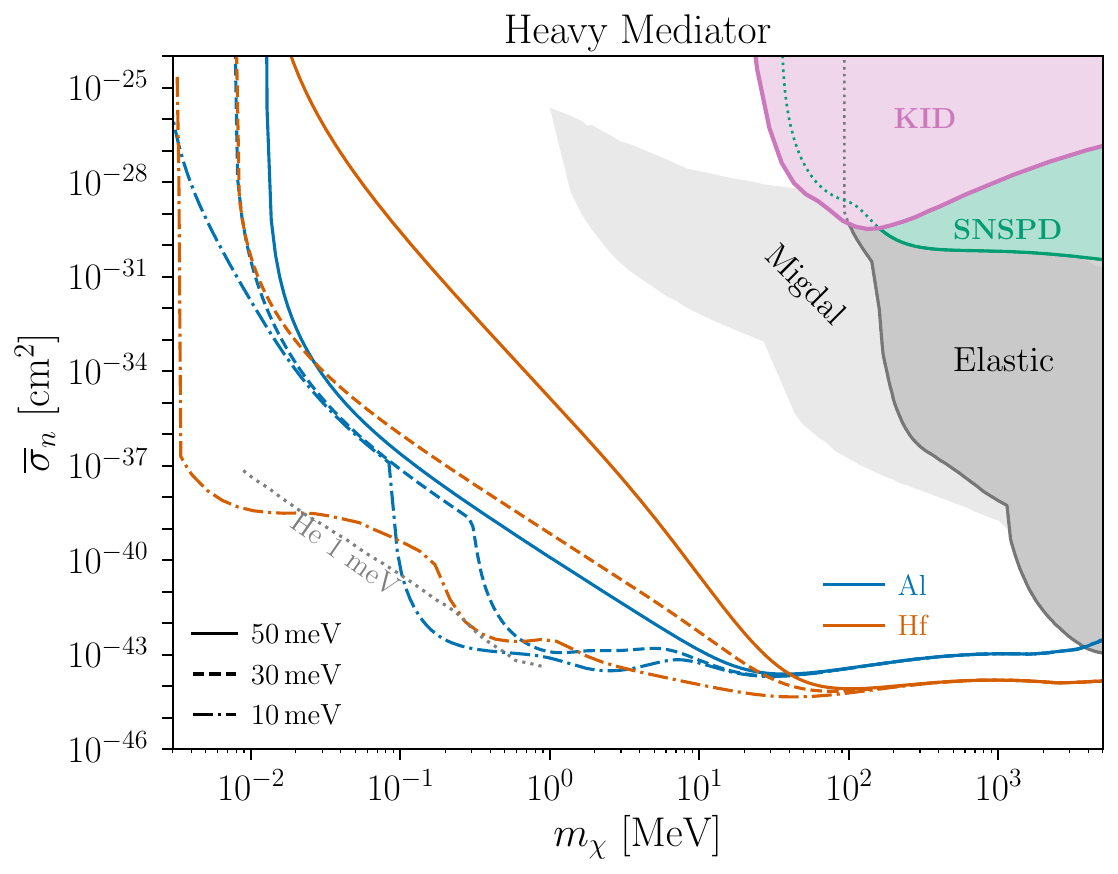}
    \caption{\textbf{Dark Matter Scattering.} Results for hadrophilic dark matter scattering with nucleons mediated by a light~(\textit{left}) or heavy (\textit{right}) scalar mediator, at the 95\% confidence level. The green and magenta shaded regions indicate the new bounds derived in this work using the SNSPD and KID data of \refscite{Hochberg:2019cyy,Hochberg:2021yud,Gao:2024irf}. The shaded dark and light gray regions indicate the strongest existing terrestrial constraints to date based on elastic interactions and the Migdal effect, respectively. These include constraints from EDELWEISS~\cite{EDELWEISS:2019vjv, EDELWEISS:2022ktt}, DarkSide-50~\cite{DarkSide-50:2022qzh, DarkSide_2023}, SuperCDMS~\cite{SuperCDMS:2020aus, SuperCDMS:2023sql}, CRESST-III~\cite{CRESST_III_2019}, SENSEI~\cite{SENSEI:2023zdf}, PandaX-4T~\cite{PandaX_4T_2023}, LUX~\cite{LUX:2018akb} and XENON1T~\cite{XENON_2020} data. The blue and orange curves show the projected sensitivity of experiments with superconducting \ce{Al} and \ce{Hf} targets, respectively, via excitations of single or multiple or phonons in the target, for a kg-yr exposure. Solid, dashed, and dot-dashed curves indicate thresholds of \qty{50}{\milli\electronvolt}, \qty{30}{\milli\electronvolt}, and \qty{10}{\milli\electronvolt}, respectively. For comparison, the dotted gray curve indicates the projected reach of a superfluid \ce{He} experiment~\cite{Knapen:2016cue} with a threshold of \qty{1}{\milli\electronvolt} and kg-yr exposure.}
    \label{fig:results}
\end{figure*} 
We have now set the stage to compute interaction rates in various targets and to determine the sensitivity of real experiments to DM parameter space. 
We use existing experimental data collected from two prototype superconducting detectors operated above ground: an SNSPD~\cite{Hochberg:2019cyy,Hochberg:2021yud} and a KID~\cite{Gao:2024irf}. These two detectors have already demonstrated extraordinarily low energy thresholds along with low dark count rates. We derive constraints by comparing the measured count rate in these experiments to the event rate implied by \cref{eq:rate-from-structure-factor}, with $S(\bb q, \omega)$ determined by \cref{eq:structure-factor-elastic,eq:structure-factor-single-phonon,eq:structure-factor-multiphonon-incoherent,eq:structure-factor-impulse-approximation} in the appropriate regimes.

The SNSPD~\cite{Hochberg:2019cyy,Hochberg:2021yud} considered here  is composed of \ce{WSi} with a mass of \qty{4.3}{\nano\gram}, and was determined to have a threshold of \qty{0.73}{\electronvolt}. A dark count rate of \qty{6e-6}{\per\second} was observed during a science run of 180 hours. For this detector, the threshold is large compared to the scale of single-phonon energies in the material, so we consider only elastic nuclear recoils. Hence our new bound is based on the interaction rate of a DM particle with the \ce{W} and \ce{Si} target nuclei within this prototype detector.

The KID~\cite{Gao:2024irf} is composed of \ce{TiN} with a fiducial volume of \qty{2e-12}{\centi\meter^3}, and a dark count rate of \qty{2e-3}{\per\second} was observed during a science run of 26 hours. We obtain the phonon spectrum in \ce{TiN} via the finite displacement method using \textsc{phonopy}~\cite{phonopy-phono3py-JPCM,phonopy-phono3py-JPSJ} with density functional theory (DFT) calculations carried out in \textsc{vasp}~\cite{Kresse1993,Kresse1994,Kresse1996,Kresse1996a} using the PBE exchange-correlation functional~\cite{perdew_generalized_1996} and projector augmented wave (PAW) pseudopotentials~\cite{Blo,Kresse1999} including 12 and 5 electrons as valence for Ti and N, respectively. We use a plane-wave energy cutoff of \qty{800}{\electronvolt}, Gamma-centered $k$-point grids with $k$-point spacing less than \qty{0.17}{\per\angstrom} and convergence criteria for the electronic self-consistent loop set to \qty{e-8}{\electronvolt}. Structure optimization using a maximum force criterion of \qty{e-3}{\electronvolt\per\angstrom} yielded the lattice constant \qty{4.249}{\angstrom}. We calculate the phonon spectrum with a 216 atom supercell on a $40 \times 40 \times 40$ $q$-point mesh. The resulting population of phonon modes is shown in the top panel of \cref{fig:phonon-spectrum}. The effective threshold\footnote{The threshold of the KID is determined by the characteristics of the noise in the device at low energies, and is subject to modeling assumptions. See \refcite{Gao:2024irf} for details.} of the device is as low as ${\sim}\qty{200}{\milli\electronvolt}$, whereas $\omega_\phonon^{\max} \approx \qty{70}{\milli\electronvolt}$. The constraints we place from existing data receive contributions from multiphonon production, and is mainly in the many-phonon impulse regime.

\Cref{fig:results} summarizes our new limits on the DM-nucleon cross section placed by the SNSPD (shaded green) and KID (shaded magenta) data, as well as projections for the sensitivity of future experiments with lower thresholds and larger exposures. Our bounds are set at the 95\% confidence level for scattering of DM via a light (\textit{left panel}) or heavy (\textit{right panel}) mediator, and we incorporate the measured count rates using the Feldman-Cousins procedure~\cite{Feldman:1997qc}. In principle, the KID constraints continue to lower masses, and fall off only as a power law for $\qty{1}{\mega\electronvolt}\lesssim m_\dm \lesssim \qty{20}{\mega\electronvolt}$, as multiphonon excitations remain kinematically available below the threshold for elastic nuclear recoils. However, we do not show constraints on cross sections above \qty{e-25}{\centi\meter^2} due to significant effects from atmospheric scattering en route to the detector. A full treatment of overburden is complicated in general, particularly for light mediators, but the cross sections we show are not typically subject to significant atmospheric scattering in surface experiments at comparable DM masses~\cite{Emken:2019tni}. For comparison, the gray-shaded regions of \cref{fig:results} depict current best existing constraints from terrestrial experiments~\cite{SuperCDMS:2020aus,SuperCDMS:2023sql,EDELWEISS:2019vjv,EDELWEISS:2022ktt,DarkSide-50:2022qzh,DarkSide_2023,CRESST_III_2019,SENSEI:2023zdf,PandaX_4T_2023,LUX:2018akb,XENON_2020}. The darker gray region in the right panel indicates direct constraints from DM-nucleon scattering, while the lighter shading indicates constraints inferred from electronic excitations via the Migdal effect~\cite{Ibe:2017yqa,Das:2024jdz}. 

The blue and orange curves in \cref{fig:results} show the projected reach for future experiments consisting of superconducting \ce{Al} and \ce{Hf} targets, respectively, with thresholds of 10, 30, and \qty{50}{\milli\electronvolt} through multiple phonon excitations in the target. Note that detectors with \ce{Hf} targets have already been demonstrated~\cite{Zobrist:2022gcj,Swimmer:2023tra}. The phonon populations we use for \ce{Al} and \ce{Hf} are shown in the middle and bottom panels of \cref{fig:phonon-spectrum}. For \ce{Al}, we use a 256 atom supercell and three electrons as valence, with all other calculation settings the same as for \ce{TiN}, resulting in a lattice constant of \SI{4.038}{\angstrom}. For \ce{Hf}, we use a 150 atom supercell and 10 electrons as valence, resulting in lattice constants $\{a, c\} = \{3.202,\,5.055\}\,\qty{}{\angstrom}$. 

These projections are made for a \qty{1}{\kilo\gram.\year} exposure, assuming no background events at a 95\% confidence level. For comparison, we also show in dotted gray the expected reach of a proposed experiments utilizing superfluid \ce{He}~\cite{Knapen:2016cue} with a \qty{1}{\milli\electronvolt} threshold and a \qty{1}{\kilo\gram.\year} exposure. The pronounced features in the low-threshold \ce{Al} and \ce{Hf} projections arise from single-phonon excitations: this channel becomes available only for thresholds $\omega_{\min} < \omega_\phonon^{\max}$. The acoustic phonons can only be produced by DM scattering for masses $m_\dm \gtrsim \qty{180}{\kilo\electronvolt}\times(10^{-5}/c_{s})\times(\omega_{\min}/\qty{10}{\milli\electronvolt})$, where $c_{s,\,\ce{Al}} \approx 2\times 10^{-5}$ and $c_{s,\,\ce{Hf}} \approx 1\times 10^{-5}$. The \ce{Hf} spectrum also includes optical phonons with energy gap $\omega_{O} \approx \qty{10}{\milli\electronvolt}$, which can be excited by $m_\dm \gtrsim \qty{3}{\kilo\electronvolt}\times(\omega_{O}/\qty{10}{\milli\electronvolt})$. For small DM masses, as long as single phonons are kinematically allowed, they dominate the rate.

\section{Discussion}
\label{sec:discussion}
We have shown how detectors geared at probing DM-electron interactions can be simultaneously sensitive to DM-nucleon interactions. This is made possible due to the electron-phonon coupling in materials, which enables energy deposits in one type of degrees of freedom to be transferred to the other. We have demonstrated the power of this approach by using published data from superconducting detectors---previously used to set world-leading limits on DM scattering with electrons---to place new constraints on DM-nucleon interactions. Computing the reach of future superconducting \ce{Al} targets into light DM parameter space, our work lays the groundwork to readily double the science that will be extracted from existing and future detectors.

This has particularly significant implications for the future of superconducting detectors. Such devices were originally proposed as a light DM scattering target almost ten years ago, by \refcite{Hochberg:2015pha}. At that time, they were introduced strictly as electron-recoil detectors, intended to be complementary to other technologies probing nuclear recoils. Our results from prototype experiments illustrate that 
there is no need for a distinction between electron-recoil and nuclear-recoil experiments: the same systems probe both interactions. Several existing and planned low-threshold electron recoil experiments are simultaneously capable of probing DM-nucleon couplings.

We stress that the prototype superconducting targets we used to place new limits on DM parameter space in this work are significantly smaller in detector mass and in exposure time than other existing detectors. Nonetheless, the low thresholds of the superconducting detectors we use already enable us to probe DM masses lower than than any previously probed by elastic DM-nucleon scattering. Indeed, efforts are already underway to achieve low thresholds with larger target masses~\cite{Cruciani:2022mbb,Temples:2024ntv}. With the future scaling of such experiments, already well-motivated by their projected sensitivity to DM-electron interactions, superconducting detectors promise to probe deep into uncharted parameter space for DM-nucleon interactions as well.

Our results from existing experiments are mostly sensitive to parameter space that is nominally probed by larger, higher-threshold semiconductor and time projection chamber experiments using the Migdal effect. In this work, we focus on the relatively simple processes that arise directly from DM-nucleon interactions: single phonon production, multiphonon production, and elastic nuclear recoils. As \cref{fig:results} demonstrates, even these processes will eventually probe much lower masses than is possible with the Migdal effect. These are, however, not the only processes that can lead to detectable events. (See, for example, \refcite{Diamond:2023fsm}, in which hadronic loops dominate for scattering via a vector mediator.) These detectors are also sensitive to the production of prompt quasiparticle pairs via an off-shell phonon. Quasiparticle pairs can offer better kinematical matching to low-mass DM, potentially extending the reach of superconducting detectors further into unconstrained parameter space. The computation of the rate will be the subject of future work~\cite{future:collective-migdal}.

\medskip

\begin{acknowledgments}
\textbf{Acknowledgments.}
We thank Noah Kurinsky for useful discussions and for comments on the manuscript. 
S.G. was supported by the US Department of Energy under the Quantum Information Science Enabled Discovery (QuantISED) for High Energy Physics grant KA2401032. Work at the Molecular Foundry was supported by the Office of Science, Office of Basic Energy Sciences, of the U.S. Department of Energy under Contract No. DE-AC02-05CH11231. 
The work of Y.H. is supported by the Israel Science Foundation (grant No. 1818/22), by the Binational Science Foundation (grants No. 2018140 and No. 2022287) and by an ERC STG grant (``Light-Dark,'' grant No. 101040019). 
The work of B.V.L. is supported by the MIT Pappalardo Fellowship. K.I. acknowledges support from the UK Engineering and Physical Sciences Research Council (EPSRC) through grant No. (EP/W028131/1). This research used resources of the National Energy Research Scientific Computing Center (NERSC), a Department of Energy Office of Science User Facility using NERSC award BESERCAP0028926. DFT computations were performed using the Sulis Tier 2 HPC
platform hosted by the Scientific Computing Research Technology Platform at the University of Warwick and funded by EPSRC grant No. EP/T022108/1 and the HPC Midlands+ consortium. Through membership of the UK’s HEC Materials Chemistry Consortium (EPSRC grant Nos.
EP/L000202, EP/R029431, EP/T022213), this work also used
ARCHER2 UK National Supercomputing Services.
This project has received funding from the European Research Council (ERC) under the European Union’s Horizon Europe research and innovation programme (grant agreement No. 101040019).  Views and opinions expressed are however those of the author(s) only and do not necessarily reflect those of the European Union. The European Union cannot be held responsible for them.
\end{acknowledgments}
\bibliography{references}

\end{document}